\documentclass[aip,reprint]{revtex4-1}
\usepackage{graphicx}
\usepackage{epstopdf}
\usepackage{amsmath}
\usepackage{amssymb}
\usepackage{siunitx}
\usepackage{natbib}
\usepackage{hyperref}
\usepackage{breakurl}

\begin{document}

\title{Vacuum-compatible photon-counting hybrid pixel detector for wide-angle X-ray scattering, X-ray diffraction and X-ray reflectometry in the tender X-ray range}

\author{D. Skroblin}
\email{dieter.skroblin@ptb.de}
\author{A. Schavkan}
\author{M. Pfl{\"{u}}ger}
\affiliation{Physikalisch-Technische Bundesanstalt (PTB), Abbestra{\ss}e 2--12, 10587 Berlin, Germany}
\author{N. Pilet}
\affiliation{DECTRIS Ltd., Taeferweg 1, 5405 Baden, Switzerland}
\author{B. L{\"{u}}thi}
\affiliation{DECTRIS Ltd., Taeferweg 1, 5405 Baden, Switzerland}
\author{M. Krumrey}
\affiliation{Physikalisch-Technische Bundesanstalt (PTB), Abbestra{\ss}e 2--12, 10587 Berlin, Germany}

\date{\today}

\begin{abstract}
A vacuum-compatible photon-counting hybrid pixel detector has been installed in the ultra-high vacuum (UHV) reflectometer of the four-crystal monochromator (FCM) beamline of the Physikalisch-Technische Bundesanstalt (PTB) at the electron storage ring BESSY II in Berlin, Germany.
The setup is based on the PILATUS3 100K module.
The detector can be used in the entire photon energy range accessible at the beamline from \SIrange{1.75}{10}{\keV}.
Complementing the already installed vacuum-compatible PILATUS 1M detector used for small-angle scattering (SAXS) and grazing incidence SAXS (GISAXS), it is possible to access larger scattering angles.
The water-cooled module is located on the goniometer arm and can be positioned from \SIrange{-90}{90}{\degree} with respect to the incoming beam at a distance of about \SI{200}{\mm} from the sample.
To perform absolute scattering experiments the linearity, homogeneity and the angular dependence of the quantum efficiency, including their relative uncertainties, have been investigated.
In addition, first results of the performance in wide-angle X-ray scattering (WAXS), X-ray diffraction (XRD) and X-ray reflectometry (XRR) are presented.
\end{abstract}


\maketitle %\maketitle must follow title, authors, abstract and \pacs

\section{Introduction}\label{sec:introduction}

The PILATUS3 detector is a silicon photon-counting hybrid pixel detector which was originally designed for protein X-ray diffraction at the Swiss Light Source~\cite{Kraft:2009}.
It has a large dynamic range of 20 bits, high count-rate capability of up to \SI{e7}{\per\s\per pixel}, high detection efficiency (\ref{subsec:quantum-efficiency}), low or zero dark noise (\ref{subsec:the-pilatus-detector-system}), and a very sharp point spread function close to the ideal of only one pixel~\cite{Kraft:2009,Donath:2013}.

An in-vacuum PILATUS 1M is part of the small-angle X-ray scattering (SAXS) setup at the four-crystal monochromator (FCM) beamline in the laboratory of the Physikalisch-Technische Bundesanstalt (PTB) and is described in detail by \citet{Wernecke:2014}.
It is successfully applied for SAXS~\cite{Garcia-Diez:2015,Gollwitzer:2016} and grazing incidence SAXS (GISAXS)~\cite{Wernecke:2014c,Pfluger:2017,Coric:2018}.
In contrast to commercially available hybrid pixel detectors which are operated in air, the in-vacuum PILATUS 1M can access lower photon energies in the tender X-ray regime covered by the FCM beamline~\cite{Wernecke:2014}.
This energy region is of high scientific importance because the absorption edges of technologically and biologically relevant elements like silicon, phosphorus, sulfur, chlorine, potassium and calcium are located here.

Larger angles with respect to the incoming beam are needed for wide angle X-ray scattering (WAXS) and X-ray diffraction (XRD) as well as for X-ray reflectometry (XRR) at lower photon energies.

Therefore, we present a new smaller photon-counting hybrid pixel detector which is installed on the $2\theta$ goniometer arm in the UHV reflectometer of the FCM beamline.
This setup was developed in collaboration with DECTRIS Ltd. and consists of a vacuum-compatible version of the PILATUS3 100K module.
It ensures that the entire accessible energy range of the beamline can be used.
However, to enable absolute scattering experiments in the tender X-ray regime an additional calibration is required.
For this reason we characterize the radiometric and geometric properties of the detector by traceable methods.
The possible applications are presented by means of the first results with WAXS, XRD and XRR\@.

\section{Instrumentation}\label{sec:instrumentation}

\subsection{The four-crystal monochromator beamline}\label{subsec:the-four-crystal-monochromator-beamline}

The FCM bending magnet beamline is part of the laboratory of PTB, the National Metrology Institute of Germany, and is located at the electron storage ring BESSY II\@.
The central optical element is an FCM equipped with four InSb($111$) and four Si($111$) crystals.
Two crystals of each kind are mounted on one wheel, and the whole monochromator consists of two wheels.
To scan the energy, the wheels counterrotate.
This monochromator design allows a fixed beam position and offers a high spectral purity and spectral resolution.
The accessible photon energy ranges from \SIrange{1.75}{10}{\keV}~\cite{Krumrey:2001}.
The energy resolving power $E_{\mathrm{ph}}/\Delta E_{\mathrm{ph}}$ is \num{e4} with an accuracy of the energy scale of \SI{0.5}{\eV}.
On the upstream side of the monochromator is a toroidal mirror which focuses the beam in the horizontal direction and collimates the radiation in the vertical direction.
Downstream of the FCM, there is a plane mirror with a pressure bender that enables two working regimes - \mbox{either} with a low divergence beam or with a focus in the vertical direction.
Attached to this beamline is a sample chamber in the shape of a cylinder with a length of \SI{700}{\mm} and a diameter of \SI{600}{\mm} and with an interlock chamber attached to the top~\cite{Fuchs:1995}.
Located in this chamber is a goniometer with 6 axes for sample movement and a detector arm with 2 axes.
The positioning accuracy of the detector arm is \SI{0.001}{\degree}~\cite{Fuchs:1995}.
The photon flux of the incident beam can be measured with a relative uncertainty of \SI{1}{\percent} by photodiodes mounted on the detector arm.
The photodiodes were calibrated against a cryogenic electrical substitution radiometer~\cite{Gerlach:2008}.

\begin{figure}
\centering
\caption{View into the reflectometer with a) PILATUS3 detector module, b) calibrated photodiodes located on the $2\theta$ arm, c) feed-through for cooling water and electronics and d) silicon drift detector for X-ray fluorescence measurements. The yellow arrows correspond to the direction of the X-ray beam.}
\label{newref}
\includegraphics[width=\columnwidth]{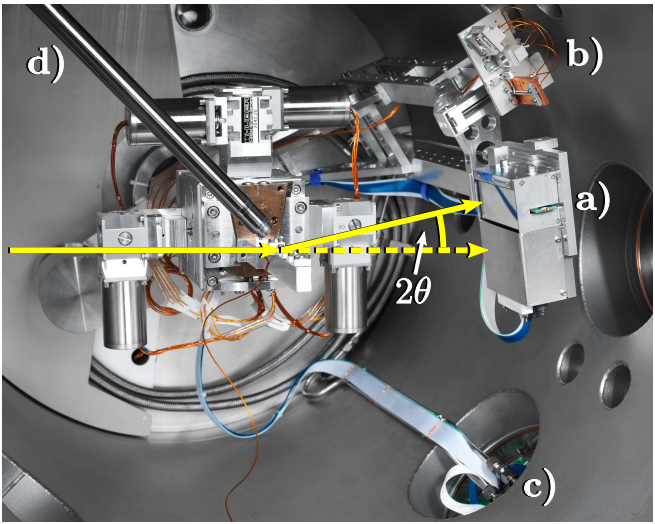}
\end{figure}

A picture of the current setup can be seen in Fig.~\ref{newref}.
The detector arm was modified and reinforced in order to carry the additional mass of about \SI{0.9}{\kg} for the PILATUS3 detector module.
A baseplate with integrated cooling water channels was designed with the possibility of mounting the long side of the detection area either horizontally or vertically.
The cooling water is supplied by a thermostat (Huber minichiller 280) with a chiller set point of \SI{5}{\degreeCelsius}.
It enters the vacuum chamber via a feed-through (Fig.~\ref{newref}c) and is directed to the back of the tank in stiff stainless steel pipes together with the electronic ribbon cable.
From here flexible tubes are loosely wrap around the goniometer and guided between a Teflon and a PEEK half disc to prevent abrasion when moving the detector arm.
Along the arm the water is again directed in stiff steel pipes to the baseplate of the detector (Fig.~\ref{newref}a).
The power input of the PILATUS3 head is about \SI{10}{\watt} and the temperature at the sensor in operation is \SIrange{7}{9}{\degreeCelsius}.

\subsection{The PILATUS3 detector system}\label{subsec:the-pilatus-detector-system}

The PILATUS3 100K (see Table~\ref{pilatus}) is a photon-counting hybrid pixel detector system.
It consists of a single pixelated silicon sensor that is bump-bonded to an array of $8 \times 2$ CMOS readout chips (ROCs).
Each pixel consists of a pn-junction with a highly doped p-electrode with dimensions of $\SI{112}{\um}\times\SI{112}{\um}$ and a pixel pitch of $\SI{172}{\um}\times\SI{172}{\um}$.
The top surface is covered by an aluminum layer to protect the sensor from visible light.
To span the gap between adjacent ROCs, the size of 2 pixels is increased to \num{3/2} of the normal pixel size so that 1 virtual pixel is inserted at the edges.
The corner of four ROCs is covered by 4 large pixels with a size \num{9/4} of a standard pixel, adding 5 virtual pixel in the corner~\cite{Donath:2013,Kraft:2009,Trueb:2012}.

Incident photons that are absorbed in the silicon sensor create electron-hole pairs.
The charge is then collected at the bottom of the pn-junction by applying a high-voltage electric field and is amplified by a charge sensitive pre-amplifier and AC coupled with a shaper to reduce electronic noise.
The analog pulse is then conducted into a comparator, producing a digital signal if the pulse amplitude exceeds the given threshold.
For each pixel, the threshold is set with a global threshold voltage and is individually trimmed with a 6-bit digital-to-analog converter.
This digital pulse then increments the 20-bit counter~\cite{Donath:2013,Kraft:2009,Trueb:2012}.

The threshold voltage corresponds to a threshold energy $E_{\mathrm{th}}$, which is the minimum energy photons must have to be detected.
Photons can create a charge that is spread over more than one pixel.
To consider this for the case where the charge is distributed over two pixels, $E_{\mathrm{th}}$ can be set to an optimal value at half of the photon energy $E_{\mathrm{ph}}$.
In this case, only one of the involved pixels will detect the photon.
In addition, the detector has the option to change between two different gain settings, namely ultra-high gain and auto-gain.
Ultra-high gain utilizes a higher amplification which yields a better detection of photons at low $E_{\mathrm{ph}}$.
The lowest adjustable threshold energy for the auto-gain is \SI{2.7}{\keV}.
Thus, below a photon energy of \SI{5.4}{\keV}, the gain mode is switched from auto-gain to ultra-high gain.
At photon energies above \SI{5.4}{\keV}, however, the detection rate is decreased compared to the auto-gain, due to the increased dead time of the longer pulses.

The dark noise of this setup is $\approx$\SI{0.2}{cph/pixel} for threshold energies above \SI{2}{\keV} and a result of cosmic radiation.
Below a threshold energy of \SI{2}{\keV} the dark noise is with $\approx$\SI{0.5}{cph/pixel} slightly higher due to dark counts in the larger pixels between the chips.

The vacuum compatible version of the PILATUS3 can be used at pressures below \SI{e-2}{\milli\bar}.
We were able to operate it reliably at pressures as low as \SI{e-7}{\milli\bar}.
We use the standard detector control unit as supplied by DECTRIS.
A socket connection (EPICS) together with the Camserver software of the manufacturer is utilized to integrate the detector into our beamline environment.
Detailed information on this can be found in the user manual available on the DECTRIS homepage \cite{:}.
The modifications to the commercially available product are the replacement of the casing with a lightweight aluminum cover and the position of the sensor area which is tilted by \SI{90}{\degree}.

\begin{table}
\caption{Technical specifications of the PILATUS3 100K system.}
\label{pilatus}
\begin{center}
\begin{ruledtabular}
\begin{tabular}{ll}
%\hline
Sensor thickness & \SI{450}{\um}\\
Pixel size & $172 \times 172\,\si{\um^2}$\\
Number of pixels (width $\times$ height) & $\num{94965}\, (= 487 \times 195)$ \\
Detection area (width $\times$ height) & $83.8 \times 33.5\,\si{\mm^2}$\\
Dynamic range & $\num{20}\,\mathrm{bits}\ (\num{1}:\num{1048576})$\\
Max. counting rate & \SI{e7}{\per\s\per pixel}\\
Energy range & \SIrange{1.75}{36}{\keV}\\
Threshold range & \SIrange{1.6}{18}{\keV}\\
Readout time & \SI{0.95}{\ms}\\
Frame rate & \SI{500}{\Hz}%\\
%\hline
\end{tabular}
\end{ruledtabular}
\end{center}
\end{table}

\section{Calibration}\label{sec:calibration}

\subsection{Quantum efficiency}\label{subsec:quantum-efficiency}

The quantum efficiency (QE) of the detector has to be known in order to obtain absolute scattering intensities.
We measure it through the direct irradiation of the detector as the ratio of the count rate to the incident photon flux.
To determine the QE precisely, the incoming flux is distributed over a $3 \times 3\,\si{\mm^2}$ area of the detector.
This is necessary in order to minimize nonlinear counting effects present at high fluxes while simultaneously having enough flux to exceed the dark current of the photodiode by several orders of magnitude.
The flux was adjusted so that the maximum pixel count rate on the detector was $\SI{2e4}{\per\s}$.
For different photon energies, sets of 3 images each with \SI{1}{\s} acquisition time are collected and averaged.
The nominal photon flux was recorded directly before and after the images were taken.
The relative standard uncertainty for the QE measurement is \SI{2}{\%} and the main contributions are statistical deviations (as determined in section~\ref{subsec:linearity}) and the flux measurement using the photodiodes.

\begin{figure}
\centering
\caption{Quantum efficiency of the detector versus photon energy.
The data represented by the red circles were collected using a fixed threshold of $E_{\mathrm{th}} = \SI{1.6}{\keV}$, whereas the common setting of $E_{\mathrm{th}} = 1/2 E_{\mathrm{ph}}$ was chosen for the blue squares.
The shaded areas around the data points indicate the corresponding standard uncertainties.}
\label{qe}
\includegraphics[width=\columnwidth]{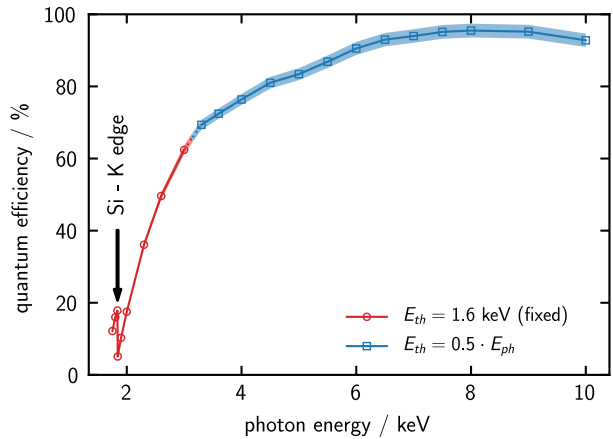}
\end{figure}

As shown in Fig.~\ref{qe}, the detector has the highest QE at \SI{8}{\keV} of about \SI{95.5}{\%}.
For lower photon energies, the efficiency decreases due to photon absorption in the non-sensitive surface layers.
For higher values, the QE is limited through the sensor thickness and is significantly higher for thicker sensors as shown by \citet{Donath:2013}.
Below \SI{3.2}{\keV}, the optimal threshold of $E_{\mathrm{th}} = 1/2 E_{\mathrm{ph}}$ can no longer be set since the lowest feasible threshold for stable operation (without breaking fuses) is \SI{1.6}{\keV}.
It was fixed at this value, which leads to a rapid drop in the QE.
For increasing ratio between $E_{\mathrm{th}}$ and $E_{\mathrm{ph}}$ the QE decreases \cite{Wernecke:2014}.
Photons that hit close to a border of a pixel might be unnoticed because of their inability to overcome the threshold due to their charge spread over multiple pixels.
The drop in the QE at \SI{1839}{\eV} (silicon-K edge) is because of photon absorption in the silicon part of the non-sensitive layer.
The dent visible in the curve at \SI{5.5}{\keV} is related to the change of the gain settings from ultra-high gain to auto-gain at \SI{5.4}{\keV}.

\subsection{Linearity}\label{subsec:linearity}

The linearity of the counting behavior is determined by investigating the total counts detected for a wide range of the incoming flux.
We exploit the detuning behavior of the monochromator to realize a variety of fluxes.
When the second wheel with crystals 3 and 4 is turned away from the optimal position, the flux drops drastically, whereas the photon energy remains almost unchanged.
To determine the detector counts at every setting, an image was collected with \SI{1}{\s} acquisition time.
The photon flux was measured with a calibrated photodiode.
For all photon energies except \SI{1.8}{\keV}, Al absorbers had to be used to attenuate the direct beam sufficiently to have it straight on the detector.
In addition, the influence of the flux distribution was evaluated for two energies by focusing the radiation.
Half of the photons can be found in 12 and 3 pixels (\SI{8}{\keV}) and in 26 and 18 pixels (\SI{5}{\keV}), for the unfocused and focused beam, respectively.

\begin{figure}
\centering
\caption{Linearity of the detector for different photon energies versus incoming total flux.
The dashed line shows the influence of the spot size change.}
\label{lin}
\includegraphics[width=\columnwidth]{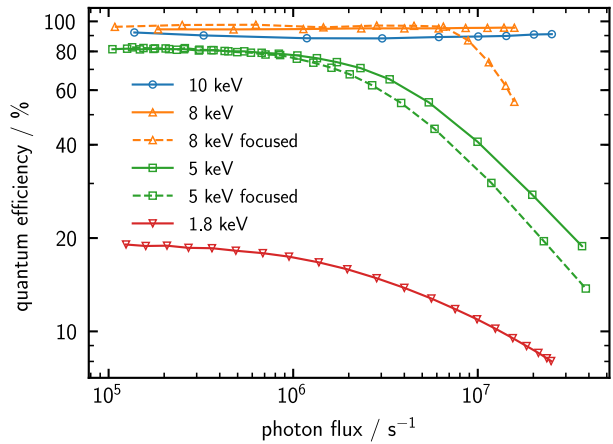}
\end{figure}

Fig.~\ref{lin} presents the result of this experiment.
For \SI{8}{\keV} and \SI{10}{\keV}, the detector shows linear behavior even for high fluxes until the top of the dynamic range (see Table~\ref{pilatus}) is reached for the single pixels.
This is clearly visible in Fig.~\ref{lin} for the focused beam at \SI{8}{\keV}, as a decrease of the efficiency at $\SI{7e6}{photon/s}$.
The relative standard deviation at these energies is below \SI{1.5}{\%}.
At \SI{5}{\keV} and \SI{1.8}{\keV}, a continuous decrease in the counts was observed.
The maximum incident rate per pixel is not reached, which indicates a real deviation from linear counting.
This is most likely due to a deviation in the detector dead time between this module and the one used for rate correction calibration.
The dead time is the input parameter used in the simple rate correction model as described in \citet{Kraft:2009}.
Calibrating this dead time is not possible with detuning the monochromator since a constant flux distribution is needed.
This can be achieved by using absorbers without altering the beam position and shape as it was done by \citet{Kraft:2009}.
For low photon energies, the strong dependence on the photon flux can result in significantly higher uncertainty contributions.
Care has to be taken with methods that can yield high count rates like X-ray diffraction.

\subsection{Homogeneity}\label{subsec:homogeneity}

To quantify the homogeneity and to characterize the applied flat-field correction, the detector was mounted on the sample goniometer.
In this position it is possible to perform line scans and 2D mappings of the detector at different photon energies.
To achieve a high spatial resolution, the detector was scanned in \SI{0.5}{\mm} steps with a pencil beam which deposited the most intensity in an area of approximately 3 by 3 pixels.
For every step an image was collected, and the total detector counts were registered.

\begin{figure}
\centering
\caption{Line scans along the long side of the detection area for different photon energies show inhomogeneities in the flatfield at the gaps between the ROCs. The gaps are depicted as gray lines.}
\label{hom}
\includegraphics[width=\columnwidth]{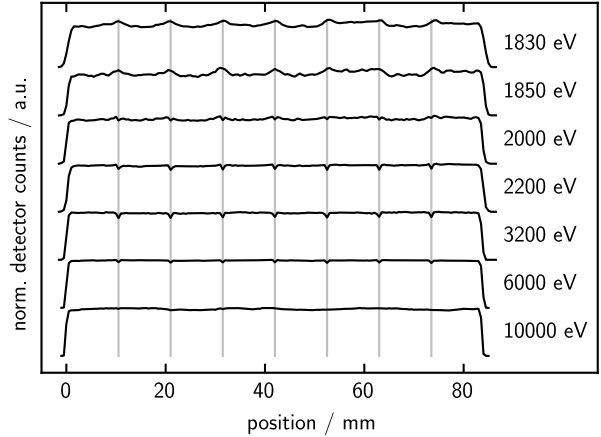}
\end{figure}

Fig.~\ref{hom} shows line scans along the long side of the detection area for photon energies from \SI{1.83}{\keV}, which is slightly below the silicon-K absorption edge, up to \SI{10}{\keV}.
The gray vertical lines denote the positions where the larger pixels span the gap between two adjacent ROCs.
The scan at \SI{10}{\keV} shows the best homogeneity.
The energy threshold is calibrated at the factory by trimming each pixel at fixed photon energies namely \SI{8}{\keV} with Cu, \SI{17.5}{\keV} with Mo and \SI{22.2}{\keV} with Ag~\cite{Donath:2013}.
In ultra-high gain mode the trim is performed at \SI{3}{\keV} with Ag and \SI{4.5}{\keV} with Ti.
For lower photon energies, there are inhomogeneities at the borders of the ROCs indicating that the linear extrapolation of the trim bits is not entirely correct.
In addition, at very low photon energies the threshold is quite close to the noise-level and thus the detector can pick up some noise counts.
These are mainly at the ROCs borders, as the larger pixels have a higher capacitance, which leads to more electronic noise.

\begin{figure}
\centering
\caption{Silicon fluorescence with $E_{\mathrm{ph}} = \SI{1.74}{\keV}$ homogeneously illuminates the detector.
The vertically integrated signal versus lines of pixels confirms the line scan measurements.
In total, 22 frames were summed, each with an exposure time of \SI{300}{\s} and $E_{\mathrm{th}} = \SI{1.6}{\keV}$.}
\label{hom2}
\includegraphics[width=\columnwidth]{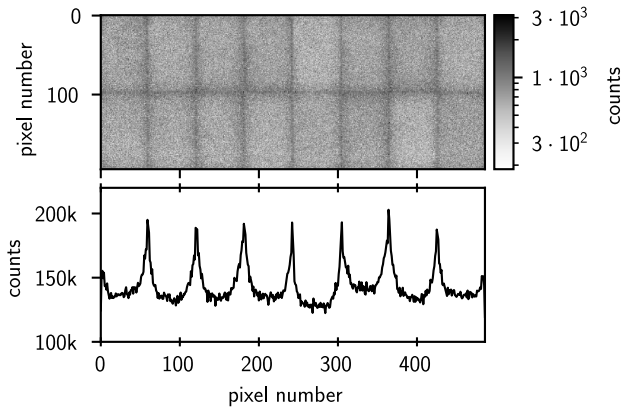}
\end{figure}

\begin{figure}
\centering
\caption{Difference between old and new flat-field for $E_{ph} = \SI{1830}{\eV}$.
For comparison the scan at \SI{10}{\keV} is shown.}
\label{ff}
\includegraphics[width=\columnwidth]{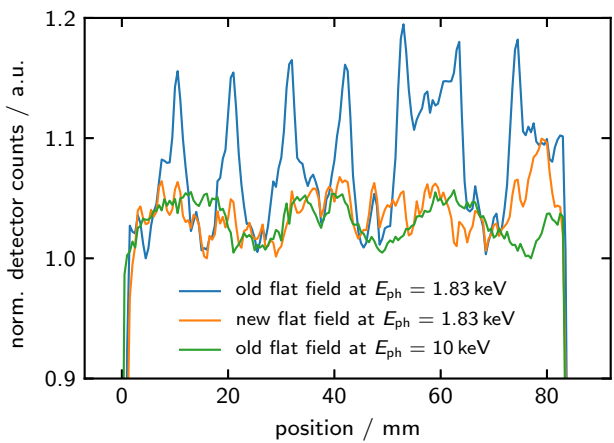}
\end{figure}

To improve the homogeneity of the detector at low photon energies, silicon fluorescence with $E_{\mathrm{ph}} = \SI{1.74}{\keV}$ was used.
For this a silicon wafer was illuminated by monochromatic light with $E_{\mathrm{ph}}=\SI{1840}{\eV}$ under grazing conditions and an image was recorded at $2\theta\approx\SI{90}{\degree}$.
In addition a solid angle correction of the intensity as implemented in pyFAI\cite{Ashiotis:2015} is applied to consider the flat detector surface (see Fig.~\ref{hom2}).
From that we extracted a flat-field which is then extrapolated to other photon energies by minimizing the line scans (~Fig.\ref{hom}).
This greatly reduces the inhomogeneity, especially at the ROCs borders, as can be seen in Fig.~\ref{ff}.
With this new flat field the relative standard deviation due to inhomogeneities remains below \SI{2}{\%} for all energies available.

\subsection{Angle of incidence}\label{subsec:angle-of-incidence}

\begin{figure}
\centering
\caption{Dependence of the detector counts on the angle of incidence (AOI) for selected photon energies.
Count rates are normalized to the number of counts for normal incidence (\ang{0}). The vertical gray line corresponds to the maximum AOI in our regular setup.}
\label{ang}
\includegraphics[width=\columnwidth]{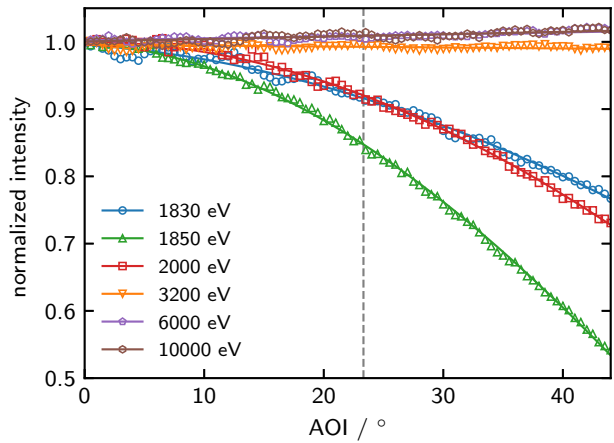}
\end{figure}

The dependence of the QE on the angle of incidence (AOI) can be used to determine the thickness of relevant layers in the detector~\cite{Krumrey:1992}.
This was investigated while the detector was mounted on the sample goniometer.
For different photon energies, the module was tilted from normal incident \SI{0}{\degree} to \SI{45}{\degree} in steps of \SI{0.5}{\degree}, and at each angle, an image was taken.
To compare the measurements, the total counts are normalized to the total counts at normal incidence.
The result is shown in Fig.~\ref{ang}.

At the tender energies used here, the angular dependence is mostly dominated by the non-sensitive layers of the detector.
Therefore, at high energies above \SI{3.2}{\keV}, there is not much change in the total counts when the angle is varied.
This is quite different for photon energies close to the silicon absorption edge.
A significant part of the non-sensitive layer is part of the silicon sensor except the thin protective aluminum layer.
When the detector is tilted, the X-rays have to travel a longer way through these inactive layers, so they are more attenuated.
That is the reason for the huge difference of the measurements performed at \SI{1830}{\eV} and \SI{1850}{\eV}.
For the lower energy which is slightly below the silicon absorption edge, the attenuation length is larger compared to the value at \SI{1850}{\eV}.
Therefore, the total counts of the tilted detector are lower for the photon energy above the edge.
The thickness of the non-sensitive part of the silicon sensor can be calculated using the following equation:
\[
t = \frac{\ln{\left(C_1/C_2\right)}}{\left(\mu_2^{-1}-\mu_1^{-1}\right)\left(\cos^{-1}{\left(\alpha\right)}-1\right)}\approx\SI{1.3}{\um},
\]
where $C_1$ and $C_2$ are the normalized counts at two different energies, $\mu_1$ and $\mu_2$ are the corresponding attenuation lengths and $\alpha$ is the AOI\@.

The maximum AOI in our regular setup is approximately \SI{23}{\degree} and is mainly given by the width of the detector and the sample to detector distance.

\section{Application examples}\label{sec:application-examples}

The PILATUS3 100K system is small and lightweight enough to be directly integrated into the UHV reflectometer of the FCM beamline.
Due to its position on the $2\theta$ arm and the small sample-to-detector distance of around \SI{200}{\mm}, it can be used for a variety of applications, two of which are presented here.
Nevertheless with this sample-to-detector distance and a pixel size of \SI{172}{\um} the angular resolution is \SI{0.05}{\degree} which limits this setup to WAXS applications.
For samples with feature sizes above \SI{10}{\nm} we use a SAXS end station with a vacuum compatible version of the PILATUS 1M detector as described by \citet{Wernecke:2014}.

\subsection{WAXS / XRD}\label{subsec:waxs-and-xrd}

\begin{figure}
\centering
\caption{Detector image of silver behenate at $E_{\mathrm{ph}} = \SI{1800}{\eV}$ for a) single position and b) stitched images of multiple positions along $2\theta$ without any data correction.
The blue rectangle matches the area of a single detector image.}
\label{AgBh}
\includegraphics[width=\columnwidth]{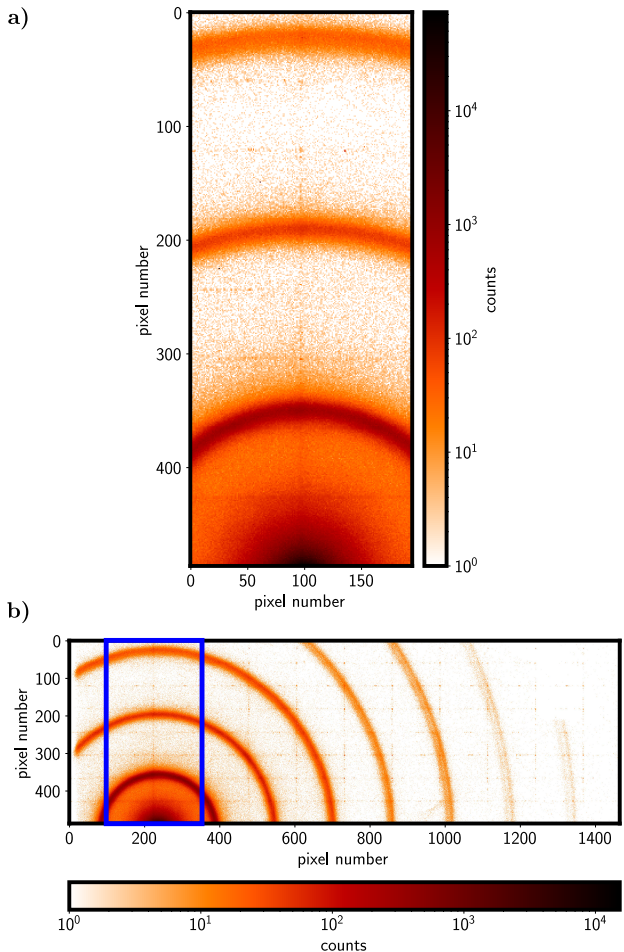}
\end{figure}

\begin{figure}
\centering
\caption{Azimuthally integrated and stitched $2\theta$ images for silver behenate (AgBh) used in SAXS, for lanthanum hexaboride (LaB6), a crystallography standard material, and for small cadmium selenide nanoparticles (CdSe NP) with an approximate diameter of $d\approx\SI{3}{\nm}$.
The vertical lines are the rough borders of the typical regions for SAXS, WAXS and XRD.
The data were collected at $E_{\mathrm{ph}} = \SI{8}{\keV}$.}
\label{WAXS}
\includegraphics[width=\columnwidth]{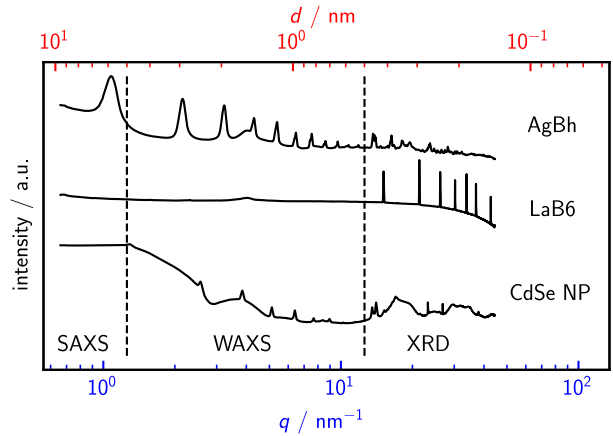}
\end{figure}

In Fig.~\ref{AgBh}, two images are presented which were collected at the photon energy of \SI{1.8}{\keV}.
They show the scattering pattern of silver behenate (AgBh) - a material that is often used in SAXS for the calibration of the $q$-axis.
Fig.~\ref{AgBh}a is a single image as it is obtained for one fixed $2\theta$ position.
With the usual SAXS setup as described in \citet{Wernecke:2014}, only the first inner ring of the material is visible.
The new system covers a wider $q$ range, enabling WAXS measurements of nano-objects.
In addition, it is possible to move the detector along $2\theta$ which extends this range even further.
A stitched image of multiple positions along $2\theta$ without further data correction is shown in Fig.~\ref{AgBh}b.
This extends the capability from WAXS to XRD experiments.
With XRD additional information on the crystal structure, the crystal size and composition of the investigated material can be gained.
Because the angle of the detector and the sample-detector distance are known, it is possible to perform azimuthal integration of the sets of images to obtain the intensity as a function of the momentum transfer $q$ or the scattering angle $2\theta$.
However, to account for the flat detector surface, it is necessary to apply a solid angle correction to each image before integrating.
Such data are displayed in Fig.~\ref{WAXS} for silver behenate and lanthanum hexaboride (LaB$_6$) (NIST SRM 660c\nocite{SRM660c:2015}), which is a common standard material for XRD\@.
Additionally, data from a measurement of cadmium selenide (CdSe) nanoparticles with a very small size ($d\approx\SI{3}{\nm}$) are shown.
These particles are synthesized via hot-injection seeded growth and are used in the synthesis of quantum dots and rods~\cite{Geissler:2017}.
The pattern in the WAXS regime originates from the electron density difference of the particle to the surroundings which corresponds to the particle shape.
The smaller peaks are due to diffraction at the self-assembled arrangement of these particles.
Similar Bragg-like features are present in silver behenate, diffraction gratings~\cite{Nygard:2010} and multiwall nanotubes~\cite{Petkov:2008}.

\subsection{XRR}\label{subsec:xrr}

Another promising application of a photon-counting hybrid pixel area detector is X-ray reflectometry (XRR).
To perform specular reflectivity, the detector is moved in $2\theta$ steps according to the $\theta$ rotation of the sample so that the specular reflection is always at the same position on the detector.
Afterwards, the intensity of the specular reflection is extracted and plotted against the grazing incidence angle $\theta$.
Compared to measurements only with a photodiode, it is possible to achieve a dynamic range for XRR experiments that is 3 orders of magnitude higher.
The resulting high $\theta$ range XRR curves enable better data modeling if, for example, it is required to show the presence of ultra-thin layers.

\begin{figure}
\centering
\caption{Combined XRR of a boron carbide and iridium multilayer at \SI{2}{\keV} using a photodiode and the PILATUS3 100K detector.}
\label{XRR}
\includegraphics[width=\columnwidth]{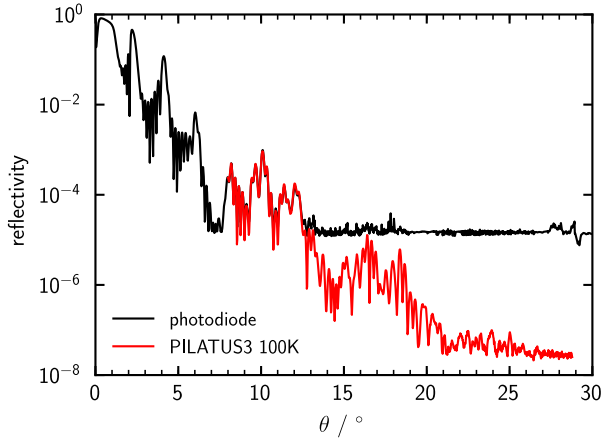}
\end{figure}

The result of such combined XRR with $E_{\mathrm{ph}} = \SI{2}{\keV}$ is shown in Fig.~\ref{XRR}.
The acquisition time for a single detector image was \SI{20}{\s}.
The lowest reflectivity measured with the photodiode is limited by the dark current.
For the PILATUS3 100K detector system, the dark count rate is close to zero thus increasing the dynamic range of the XRR measurement.
The combination of photodiode and detector is needed as the intensity of the specular reflection is too high for the detector at small angles.
The detector can be used only above a particular $\theta$ to ensure a linear response (see~\ref{subsec:linearity}).
It is also possible to replace the step-by-step scanning by a continuous data acquisition mode as described by \citet{Mocuta:2018} since the readout time of the PILATUS3 100K system is fast enough at \SI{0.95}{\ms}.
This would offer very fast XRR scans that allow dynamics and evolution phenomena to be studied.
There would however be higher uncertainties for the reflectivity.

\section{Conclusion}\label{sec:conclusion}

A vacuum-compatible version of the PILATUS3 100K detector has been successfully installed in the UHV reflectometer of the PTB four-crystal monochromator beamline at BESSY II\@.
It is mounted on the $2\theta$ arm and can be used for WAXS, XRD and XRR measurements in the entire energy range of the beamline from \SIrange{1.75}{10}{\keV}.

The linearity, homogeneity and the angular dependence of the quantum efficiency, including their relative uncertainties, were investigated.
The calibration described in this paper is mandatory for performing absolute measurements in the tender X-ray regime.
The leading uncertainty is due to inhomogeneity between the ROCs and their borders.
When using the additional flatfield collected at \SI{1.74}{\keV}, it is possible to reduce these deviations.
After this correction the relative standard deviation due to inhomogeneities remains below \SI{2}{\percent} for all energies available.
The nonlinear counting behavior has an influence on measurements at low $E_{\mathrm{ph}}$ but can be controlled by reducing the incoming photon flux.
The uncertainty contribution from the angle of incidence only has to be considered for photon energies close to the silicon absorption edge.

\begin{acknowledgments}
L. Cibik and M. Luttkus of PTB are kindly acknowledged for the technical implementation of the detector system.
S. Schreiber is especially acknowledged for the design of the improved $2\theta$ arm and the baseplate with an integrated cooling water supply.
F. Weigert and D. Gei{\ss}ler of the Federal Institute for Materials Research and Testing (BAM) are acknowledged for providing the CdSe nanoparticle sample.
S. Massahi from the Danish Technical University (DTU) is acknowledged for supplying the Ir and B$_4$C multilayer system.
\end{acknowledgments}

\bibliography{pilatus_arxiv}

\end{document}